# A novel experimental bench dedicated to the accurate radiative analysis of photoreactors: the case study of CdS catalyzed hydrogen production from sacrificial donors

G. Dahi[(1,2)] A. Eskandari[(1,2)], J. Dauchet[(3,2)], F. Gros[(3,2)], M. Roudet[(3,2)], J. F. Cornet[(3,2)]

(1) Université Clermont Auvergne, Université Blaise Pascal, Institut Pascal, BP 10448, F-63000 CLERMONT-FERRAND, France
(2) CNRS, UMR 6602, IP, F-63178 AUBIERE, FRANCE
(3) Université Clermont Auvergne, ENSCCF, Institut Pascal, BP 10448, F-63000 CLERMONT-FERRAND, FRANCE
Tel : 00 33 4 73 40 53 17
Email : fabrice.gros@univ-bpclermont.fr

**Abstract**

This article is dedicated to the presentation of a novel experimental bench designed to study the photoproduction of $H_2$. It is composed of three main parts: a light source, a fully equipped flat torus reactor and the related analytical system. The reactor hydrodynamic behaviour has been carefully examined and it can be considered as perfectly mixed. The photon flux density is accurately known thanks to reconciled quantum sensor and actinometry experiments. The incident photon direction is perpendicular to the reactor windows; in such a configuration the radiative transfer description may be properly approximated as a one dimensional problem in Cartesian geometry. Based on accurate pressure measurement in the gas tight photoreactor, the production rates of $H_2$ (using CdS particles in association with sulphide and sulfite ions as hole scavengers) are easily and trustingly obtained.

First estimations of apparent quantum yield have proven to be dependent on mean volumetric rate of radiant light energy absorbed hence demonstrating the need for the use of a radiative transfer approach to understand the observed phenomena and for the proper formulation of the thermo-kinetic coupling.

**Keywords:**

Experimental bench, photoreactor, $H_2$, photocatalysis, overall quantum yield

**Highlights:**

A novel and accurate experimental bench for the study of photoreactions is presented.

Complete characterization of the bench has been done: RTD, photon flux density…

$H_2$ production rates are accurately determined by pressure measurements.

Experimental results show a quantum yield fall with incident photon flux density rise.

**1 Introduction**

Given the scarcity of fossil resources and the constant augmentation of energy demand (between 25 and 30 TW predicted in 2050, [1]), Humanity will soon have to cope with an energetic issue together with a global warming problem due to $CO_2$ emissions. Hence, use of renewable resources to overcome these dramatic issues is compulsory. In particular resorting to solar energy is clearly evident as it is abundant. However this resource is fluctuating in nature (day/night alternation, season changing), it is thus necessary to convert it not only into electricity but also into storable chemical energy carriers synthesized from water and eventually $CO_2$.

One of the most promising (and most studied) solar fuel is undoubtedly the hydrogen, $H_2$, (known to release high quantity of energy at combustion without $CO_2$ generation [2]). Its generation is often called "artificial photosynthesis" [3], expression also utilized for the synthesis of other fuels like $CH_4$, $CH_3OH$ [4]...

Nevertheless, future $H_2$ massive renewable production needs to eliminate a double bottleneck. On the one hand, at a fundamental level, the synthesis of cheap and efficient photocatalysts (possibly designed with bio-inspired approach and mimicking natural photosynthesis) should be obtained, which is often called the Holy Grail of chemistry. On the other hand, considering the engineering science (our field of interest), one must succeed in developing large scale optimized and safe processes for the production of $H_2$. Provided breakthroughs are obtained at both levels unlocking those scientific barriers, thermodynamic efficiencies far higher than 10% could be reached. Such a value is considered as a minimum threshold of interest for the production of solar energy carriers due to competition for footprint.

Among the possibilities to produce $H_2$ from solar energy, the implementation of a solid photocatalyst suspended in slurry reactor would be a good technical solution. Hence numerous kinds of photoreactors are described in the literature, and used for catalyst efficiency characterisation or $H_2$ production. They are mainly operating in batch mode [5] presenting planar [6] or annular ([7], [8]) geometries. An external liquid circulation loop can also be used with planar photoreactor [9], [10], falling film [11] or flat plate [12]... Accurate description or modelling of photocatalytic processes requires addressing photonic phase balance (photon flux density, transmittance, radiant power absorbed …) or radiative transfer within the reactor. This crucial radiative aspect is gaining increasing interest in the community as can be seen in the work of Alfano and Cassano's team ([13] for example) or De Lasa's group [14]). We intend to place our approach and scientific work in this context to get preliminary information prior to establishing thermo-kinetic coupling relations between photoproduction rates and the radiation field in the photoreactor [15].

The aim of this article is therefore to present a novel experimental bench dedicated to accurate radiative analysis of photoreactor and its implementation in hydrogen production from sacrificial donors and CdS using a simple and well known photoreaction [16]. The first section will be devoted to the presentation and the characterisation of the bench and its three main components i.e. the photoreactor, its light source and the analytical devices for chemical reaction rate measurements. The second part will be focused on the model photoreaction. Experimental result analysis will be carried out according to an approach widespread in the literature but also by taking carefully into account energy balance on the photonic phase and radiative related aspects to gain insights into photochemical process.

## 2 Experimental devices and characterisations

### 2.1 Torus reactor

The photoreactor used in this study is a flat torus reactor intrapolated on the basis of a pilot plant used as photobioreactor [17] or as a chemical reactor with fully-controlled hydrodynamic conditions [18]. This geometry presents two translucent faces made of glass (or quartz for UV irradiation): the first in front of the lighting source, the second at the rear of the reactor enabling the measurement of photon flux densities (PFD) (cf. Fig.1). In such a design the radiative transfer modelling can be correctly approximated as a one-dimension problem.

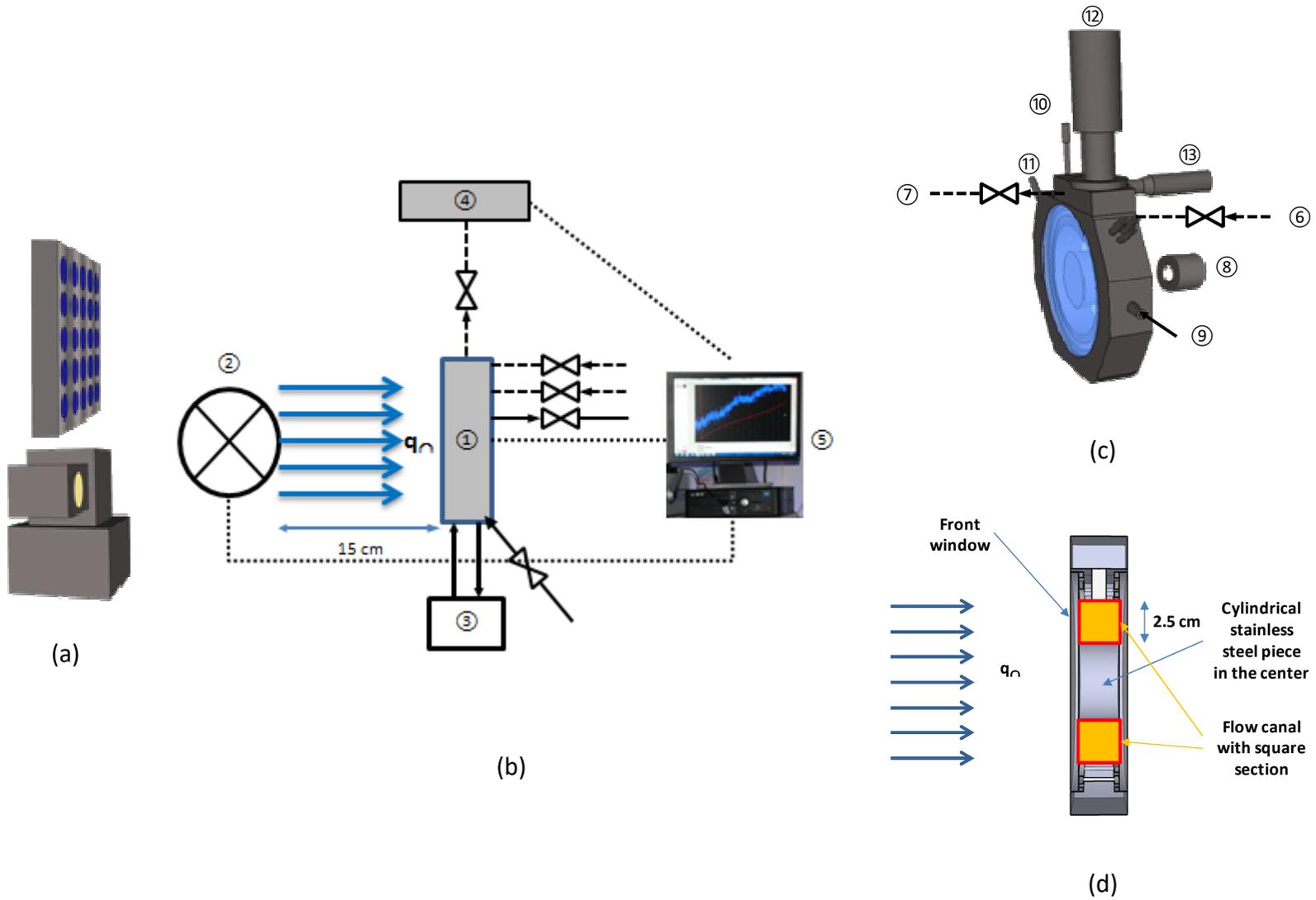

Fig. 1. (a) Scheme of the light sources LED panel or solar simulator

(b) Schematic diagram of the characterisation bench (① Flat torus reactor, ② light source, ③ thermostatic bath, ④ gas chromatograph, ⑤ control unit)

(c) Focus on the reactor and its different parts (⑥ and ⑦ pneumatic valves for gas phase handling, ⑧ quantum sensor for transmittance measurements, ⑨ entry of coolant, ⑩ pH probe, ⑪ temperature sensor, ⑫ stirring motor, ⑬ pressure sensor)

(d) Horizontal cross section of the reactor presenting the square flow canal

The torus geometry is obtained by the integration of cylindrical piece of metal in the centre of the reactor. Manufactured from a piece of 316L stainless steel, it can be seen in Fig. 1. The flow canal displays a 2.5 cm sided square section and the illuminated reactor surface, $S_{light}$, it is equal to 59 cm².

Different lateral inlets/outlets permit the loading/draining of the reactor with liquid or gas in order to, for example, operate in continuous modes for the liquid or the gaseous phase. The insertion of probes is also possible.

A liquid circulation circuit is machined in the stainless steel piece, in parallel to the flow canal. It enables the flow of a coolant fluid (water for example) using thermostatic bath/circulator (Lauda eco RE 415); a custom built RTD Pt100 sensor (TCDirect) is laterally positioned in order to set the reactor temperature at $25.0 \pm 0.1°C$. A lid is located at the top of the reactor; it offers additional possibilities for inlets or outlets and a support for the mixing device. It is constituted of a micromotor 24V/DC (Minisprint Magnetic Stirrer made by Premex Reactor ag) using magnetic coupling technology that ensures gas tightness. The motor rotation speed is controlled with a digital speed display DZA-612Z. Connected to the motor by a shaft, an helical impeller makes the liquid rotate in the annular space, suspending the photocatalyst particles. Above all this torus configuration has been chosen as it is the only way to effectively obtain homogeneous catalyst suspensions in the whole reactor.

Located in the headspace of the reactor, a Keller PA 33 X pressure sensor linked to a K107 converter is used for the accurate detection of gas production via a pressure increase (relative accuracy $\pm 0.05\%$). This peculiar sensing method will be developed later in this article considering the interface between gas and liquid phases and transfer phenomenon (see also appendix). In standard operating conditions, the reactor contains a volume of liquid, $V_L$, equal to 155 mL (evaluated by mass measurement of the loaded liquid amount), a headspace of volume $V_G$ equal to 35.2 mL (located at the top of the reactor and estimated by difference between $V_L$ and the total volume of the reactor). Furthermore, the reactor includes a small unlighted fraction at the top, $f_d$, equal to 5%.

Two pneumatic valves (Carten CMDA250) at the inlet and outlet of the reactor allow closing and opening reactor for inerting or purging. For each experiment, leak tests have been performed with hydrogen, and measured leakage rates are proved negligible.

A micro gas chromatograph (Agilent 3000A Micro GC equipped with a 5 Å molecular sieve and argon as gas carrier) completes the system; it is positioned at the reactor gas outlet to analyse online the composition of the reactor gaseous phase.

## 2.2 Residence time distribution set up and results

A residence time distribution (RTD) study was carried out in order to determine the hydrodynamic behaviour of the flat torus reactor. The RTD measurements were performed by a conductimetric method (i.e. measurement of the liquid electrical conductivity after the injection of a concentrated conducting electrolyte), using a two-pole conductivity cell (CDC749, Radiometer Analytical) located in the reactor, a CDM210 conductimeter (Radiometer Analytical) linked to a data acquisition unit (34970A, Hewlett Packard) and a computer. In such a configuration, the maximal acquisition frequency is equal to 2 Hz. Experiments were done out in batch conditions, on the liquid phase, with a pulse of 0.5 mL NaCl solution (2 g.L$^{-1}$) injected in the upper part of the reactor.

Once the ionic tracer pulse injected in the liquid, it is caused to circulate due to the pumping action of the impeller on the liquid. The conductivity signal measured by the probe corresponds to the superimposition of a periodic function (with a period corresponding to the circulation time, $t_c$) on an exponential decrease due to the mixing [18] (see Fig. 2).

The relative height of peaks is due to local high concentration of tracer before its homogenization, and their number depend on axial dispersion ($D_{ax}$) in the reactor. Better the reactor is mixed during one circulation; stronger is the decay between the consecutive peaks and consequently their number decreases. At mixing time $t_m$, the tracer is evenly distributed throughout the reactor and its concentration is constant and equal to $C_\infty$. Fig. 2 presents results of a typical RTD experiment.

In such experimental configuration, according to Benkhelifa et al. [18], an axial dispersed plug flow with total recirculation model can be chosen. In this model the circulation region is described by a tube wherein a certain amount of dispersion happens and complete recirculation of the flowing fluid occurs [19].

Voncken et al. [19] proposed a mathematical solution, later modified by Takao et al. [20] (Eq. 1) to this model for the dimensionless concentration evolution.

$$\frac{C(\Theta)}{C_\infty} = \frac{1}{2}\sqrt{\frac{Pe}{\pi\Theta}}\sum_{j=1}^{\infty}\exp\left(-\frac{Pe\left(j+z^{\cdot}-\Theta\right)^2}{4\Theta}\right) \tag{1}$$

Hence after adimensionnalisation of experimental results, they were compared with theoretical evolution as can be seen in Fig.2 (using dimensionless variables $\Theta$ (for time) in abscissa and the ratio $C(\Theta)/C_\infty$ in ordinate).

Peclet number, Pe, and circulation time, $t_c$, were identified by comparison with experimental results via a simplex method. The agreement between model and experiment is good, considering the limitation of our acquisition system. Thus for a 400 rpm rotation speed, a Peclet number equal to 8 was identified. For higher rotation speed, *i.e.* 1100 rpm, corresponding to our operating conditions in hydrogen production experiments, oscillations cannot be detected. In less than one second the tracer concentration is quasi homogeneous in the reactor.

So, the liquid phase in our flat torus reactor thus presents a high axial dispersion and a low circulation time (less than 1 s) under stirring. It can be considered to a certain extent as perfectly mixed. This crucial assumption will simplify the treatment in hydrogen production and radiative transfer analysis.

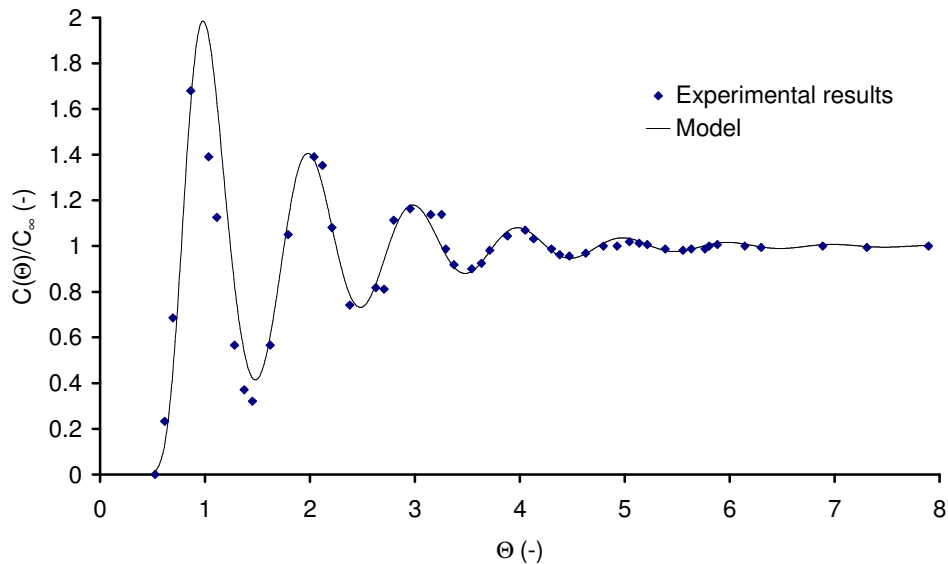

Fig. 2. Example of experimental results and model (Eq. 1) in RTD experiment (400 rpm)

## 2.3 Light source characterisation

The LED panel is fabricated by Sybilux and is composed of 25 LED (Royal blue D42180, Seoul Semiconductor) equipped with lenses. It provides a blue light with an emission maximum at 457 nm wavelength (see Fig. 3.), on a 12.5×12.5 cm surface. Such a quasi-monochromatic light source has been adopted for ease of use and post treatment. Indeed the photon flux density can be easily and accurately controlled via an USB DMX controller and the Easy Stand Alone software (Nicolaudie), including 256 different setting positions to modify the electric power supplied to the LED and thus the photon flux density. For further work, we intend to use a 450 W solar simulator (model nr 91195, Oriel) with neutral filters (Oriel series).

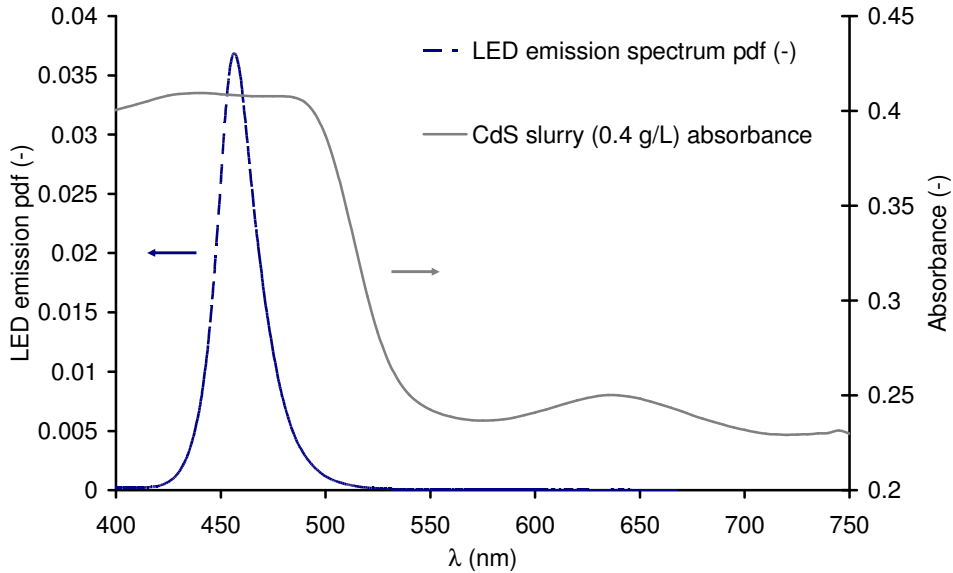

Fig. 3. LED emission probability density function (left ordinate axis) and CdS slurry

absorbance (right ordinate axis) spectra

In order to get reliable data on photochemical process, the hemispherical incident photon flux densities (PFD), $q_{\cap,\,in}$ (expressed in $\mu mol.m^{-2}.s^{-1}$) was determined via two ways. Firstly it was estimated by a physical method using a LiCOR quantum sensor (LI-190) connected to a LI-189 display. As a preliminary work, $q_{\cap}$ was measured on 56 points on a parallel planes at different distances from the LED panel. The incident photon flux densities in the region of interest (circulation canal of liquid) were also evaluated using a 16 hole mask behind which the quantum sensor was positioned. This physical method permits the determination of the mean value and the geometric standard deviation of the photon flux densities in the region of interest for a given software settings.

The second method is the actinometry where a chemical actinometer undergoes a light-induced reaction (at a certain wavelength, $\lambda$) for which the quantum yield, $\Phi_\lambda$, is accurately known. Measuring the reaction rate allows the direct estimation of a mean integrated photon flux density (instead of averaging measures obtained in several positions in the sensor method) or the calculation of the absorbed photon flux [21].

### 2.3.1. Physical method

A preliminary work has been carried out to set the optimal working distance between the LED panel and the reactor. We concluded from that study that a short distance (few centimetres) implies a non homogeneous photon flux density due to point source of the LED, and a too long gap (20-50 cm) enlarges needlessly the process. Hence the work distance was set to 15 cm. Besides, via this preliminary work, the light source could be assumed as quasi-collimated in the modelling procedure.

As previously explained, the mean photon flux densities and their standard deviations were estimated in the region of interest via 16 measurement points. In the same way as mean flux density, standard deviation increases with the Easy Stand Alone software settings, the relative uncertainty remains constant and equal to 11%, it may appear high but this finding is mainly due to edge effect, the LED panel having only 5 LEDs in height and width.

### 2.3.2. Chemical method

The Reinecke's salt $Cr(NH_3)_2(SCN)_4^-$ (supplied by Sigma Aldrich) has been chosen in this study as the actinometer as proposed by Wegner and Adamson [22] and later modified by Cornet et al. [23]. Its decomposition Eq. (2), with a constant quantum yield (equal to 0.31 in the LED emission spectrum), generates thiocyanate ions whose concentration can be determined spectrophotometrically, according to the protocol presented in [23].

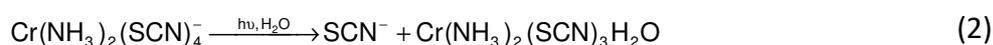

$$Cr(NH_3)_2(SCN)_4^- \xrightarrow{h\nu, H_2O} SCN^- + Cr(NH_3)_2(SCN)_3H_2O \qquad (2)$$

In this work, the analytical developments used for $q_{\cap, in}$ assessment were improved by working on longer than usual reaction durations and taking into account the non negligible proportion of photons absorbed by the by-product as explained in [24]. The photon flux density, $q_{\cap, in}$, was identified by comparing experimental results and model, the standard deviation on the identified value was estimated to 4%. A more detailed explanation of these new developments for Reinecke salt actinometry will be the object of another publication [25].

Fig. 4. represents the parity chart between the photon flux density estimated by the two methods (physical and chemical) in the whole operating range of the LED panel.

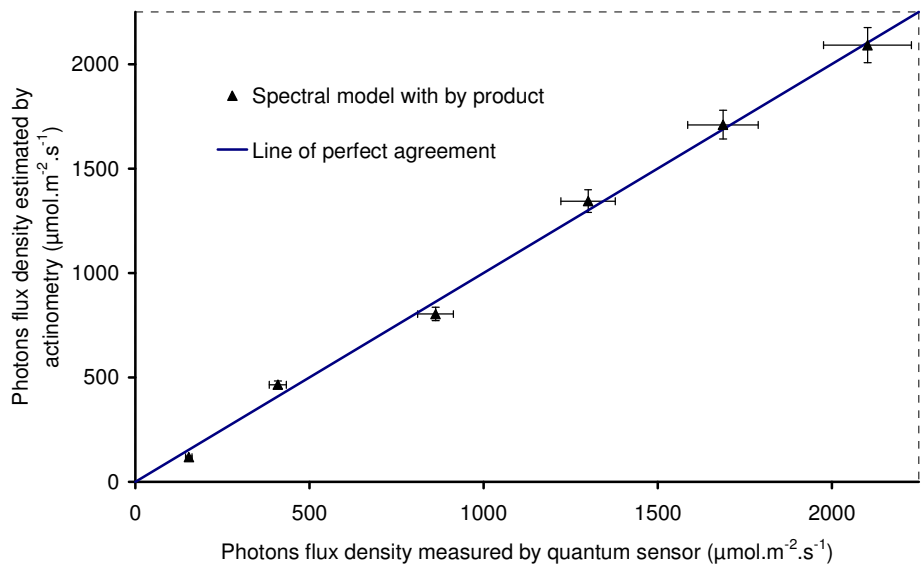

Fig. 4. Parity chart of identified values of hemispherical photon flux density, $q_{\cap, in}$, obtained in actinometry and mean values measured by quantum sensor

The results obtained by the two methods are in very good agreement. The photons flux densities being accurately known, this step allowed us to determine with confidence the boundary conditions of our system, *i.e.* the hemispherical photon flux density, $q_{\cap, in}$.

### 2.3.3. Estimation of exiting photon flux density and slurry transmittance

As already explained, the particular feature of our torus reactor is to include a rear glass window (see Fig. (1)). This enables the measurement (with the LiCOR quantum sensor) of the mean uncorrected photon flux density outgoing the reactor, $q_{\cap, \text{ out uncorrected}}$, over the entire window area with the same 16 hole mask as for $q_{\cap, \text{ in}}$ estimation. The correction of rear glass absorptance has been taken into account, using laws of geometrical optics for the quasi-collimated case, to correctly estimate the mean photon flux density outgoing the slurry, $q_{\cap, \text{ out}}$ ($q_{\cap, \text{ out}} = q_{\cap, \text{ out uncorrected}} / T_{\text{glass}}$). A mean slurry transmittance, $\tau$, could be then estimated (Eq. 3), from which energy balance on the photonic phase and absorptance calculation can be performed.

$$\tau = \frac{q_{\cap, \text{out}}}{q_{\cap, \text{in}}} \tag{3}$$

### 2.4. Reaction rate measurement

In this work, the estimations of reaction rates are performed by an accurate pressure measurement as already used by [10]. This method achieves an accuracy and a reliability far higher [26] than those obtained by gas balance using specific analyser or gas chromatography [5] and even more than those estimated with volumetric method, as often used in estimation of hydrogen production (see [27] for example). At a constant temperature, known gas and liquid volumes and considering that equilibrium is reached between gas and liquid phases (see appendix), the mean production rate for a given gas can be deduced from pressure variation inside the reactor associated with a mass balance on $H_2$ component:

$$<r_{\text{gas}}> = \left[ \frac{V_G}{V_L RT} + \frac{1}{H_{\text{gas}}(T)} \right] \frac{dP}{dt} \tag{4}$$

At a constant temperature and at a steady photon flux density, pressure increases linearly with time during $H_2$ evolution. In this case, it is easy to estimate $\left\langle r_{H_2} \right\rangle$ from the slope of the recorded pressure straight line. Liquid/gas transfer phenomenon does not appear in (Eq. 4); however for sake of scientific rigour this process should be taken into consideration to properly establish the experimental procedure in estimating production rate. A more complete modelling of pressure evolution in our flat torus reactor is presented in the appendix section of this publication.

## 3. Dihydrogen production reactions

### 3.1. Model system

In this first study devoted to the presentation of our experimental bench, we intend to work with a simple and well-known photoreaction for ease of modelling and validation of a treatment procedure. Among all the possible systems for producing $H_2$ that could be found in the literature using a photocatalyst dispersed in a solution of reducing agents acting as hole scavengers, one can find the photoproduction reaction of dihydrogen from sulphite and sulphide solution (acting as hole scavengers) in the presence of a dispersed catalyst composed of CdS. This reaction has been selected because it is simple to implement and well known in literature ([16], [28] [29] and [30] for examples). The main photoreactions in these systems are:

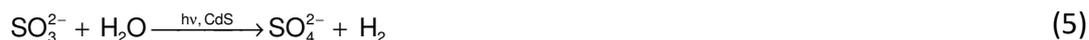
$$SO_3^{2-} + H_2O \xrightarrow{h\nu, CdS} SO_4^{2-} + H_2 \qquad (5)$$

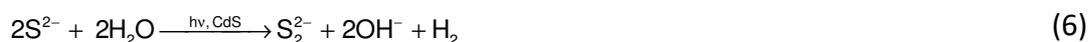
$$2S^{2-} + 2H_2O \xrightarrow{h\nu, CdS} S_2^{2-} + 2OH^- + H_2 \qquad (6)$$

Besides it should be noted that reaction (5) is part of a photo-thermochemical cycle developed in [31], where water is splitted in $O_2$ and $H_2$; reaction (6) can also be used in Clauss cycle for flue gas treatment [32] creating other opportunities to these simple reactions. They were thus implemented in the photoreactor to precisely determine dihydrogen production rate, the apparent quantum yield, $\Phi$, and their evolutions versus radiative quantities, $q_{\cap, in}$ and $<\mathcal{A}>$ (see part 4.2).

## 3.2. Reaction and implementation in reactor

Suspensions are prepared using outgassed deionised water (Milli-Q water, Millipore), sodium sulphite (Sigma Aldrich) and sodium sulphide salts (Sigma Aldrich), sodium dodecylsulphate, SDS, (Acros Organics) as dispersant agent and cadmium sulphide (supplied by Acros Organics) with 99.999% purity. The mean diameter of CdS particles is estimated to be 2.5 μm after scanning electronic microscopy analysis.

In a typical suspension, the respective concentration of sulphite, sulphide ions are 0.5 and 0.25 M. The mass concentration of cadmium sulphide particles is ranging between 0.02 and 0.4 g.L$^{-1}$, SDS is added in a mass ratio 5/1 compared to CdS concentration to ensure an homogeneous slurry with individual catalyst particle. An example of absorption spectrum for a suspension with a 0.4 g.L$^{-1}$ CdS concentration can be seen in Fig. 3. The band gap could be estimated to a value of 2.4 eV in agreement with literature data ([33] for example). The CdS particles are thus able to absorb photons with a wavelength in the range [300-520 nm] like the ones emitted by the LED panel.

The slurry is magnetically stirred and sonicated in the dark with an IKA U50 Control sonicating needle (set at full amplitude and power) for one hour to disperse aggregates into singular particles and then introduced in the reactor. The thermostatic bath is then switched on to control the inner reactor temperature at a $25 \pm 0.1°C$ value during all the following steps.

The reactor windows are covered with protections to keep it in obscurity. In order to eliminate air from the reactor, it is inerted with a flow of argon (Messer, 99.999% purity) for one hour. Considering the characteristics of the system ($k_L a$, $V_G$, $V_L$, argon flow rate…), this duration is sufficient to remove any $O_2$ or $N_2$ remaining traces. Argon being the carrier gas for chromatography too, the only detected gas would be the produced ones in the reactor. After the blanketing stage, the reactor is closed; covers are removed letting photons radiate the slurry.

### 3.3. Measurement of $H_2$ mean production rate $< r_{H_2} >$

The measurement of $< r_{H_2} >$ is performed by pressure measurement (see section 2.4, Eq. 4 and appendix). The Henry constant for dihydrogen in our solution is calculated to take into account the influence of ions on the solubility (salting out effect) [34] compared to pure water [35]. Its value is estimated to be equal to $H_{H_2}$ = 171033 $m^3.Pa.mol^{-1}$.

### 4. Results and discussions

### 4.1 Kinetic description of the $H_2$ photo-production

Once the reactor is illuminated, the inside pressure increases due to hydrogen production as can be seen in Fig. 5. This pressure rise starts with a transient state due to desorption of hydrogen generated, as explained and modelled in Appendix.

After the transient state is finished, the pressure increases linearly with time, a pseudo steady state phase is reached. According to Eq. 4, the volumetric rate of hydrogen production can then be considered as constant.

If pressure attains a high value (around 10000 Pa on Fig. 5), the outlet valve is opened to purge the reactor gas phase, the exiting gas are then directed to the chromatograph for analyses. This implies a sudden pressure drop in the reactor as can be seen in Fig. 5. (at 102 and 165 minutes for example). The outlet valve is closed right after that and the incident photon flux density is changed.

Immediately, the pressure increases quickly due to dissolved $H_2$ desorption. Afterwards another pseudo steady state can be observed. These closing and opening of the outlet valves are repeated as many times as necessary.

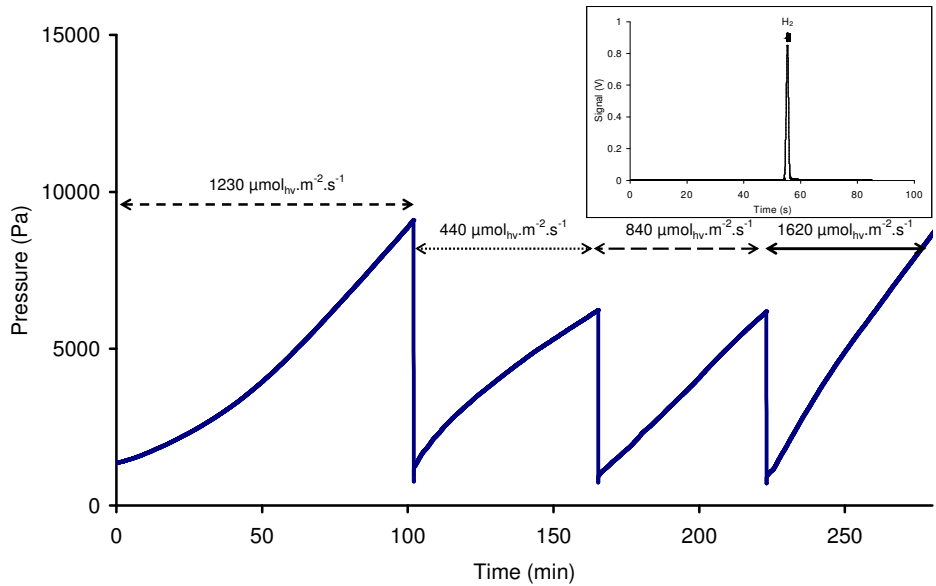

Fig. 5. Typical pressure evolution inside reactor for different photon flux densities (CdS concentration 0.1 g.L$^{-1}$) with releases of produced $H_2$; Inset hydrogen peak identified by gas chromatography

In Fig. 5, it is also possible to notice the influence of photon flux density on the pressure increases (and on the mean volumetric rate of $H_2$ production). Indeed, the higher the incident PFD, the higher the pressure rate. This confirms the photo-induced nature of the activation of the photocatalytic process of hydrogen generation by CdS particles, with the participation of photo-induced electrical charges (electrons and holes) to the reaction mechanism [36].

As a proof of hydrogen production, gas chromatography analyses have been carried out on the produced gas leaving the reactor. In Fig. 5, the inset indicates that the only detected gas is $H_2$, traces of argon that could be present in the reactor isn't detected as it is the gas carrier. The area of the peak related to $H_2$ also increases with time a sign that hydrogen concentration in the headspace is augmenting.

To ensure the repeatability of the results between the beginning and the end of an experiment, the reactor was finally exposed to the same photon flux density as at the beginning. The measured pressure increases are the same, although the solution composition slightly changed (maximum 2% conversion rate, estimated on the total pressure rise).

The influence of catalyst concentration has been studied too in the range 0.02 to 0.4 g.L$^{-1}$. A standard mode of representation of the results would be to plot the mean production rate versus hemispherical photon flux densities, $q_{\cap, in}$, for concentration series as can be seen in Fig.6a. The other type of graph, frequently seen in literature, is the plot of mean production rate versus concentrations for hemispherical photon flux density series (Fig. 6b)).

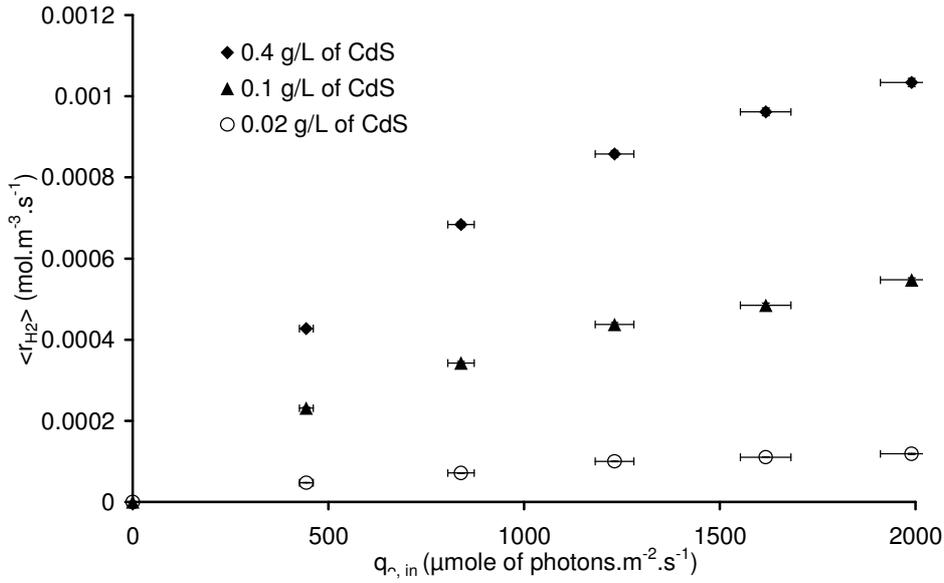

Fig. 6a. Mean production rate of hydrogen versus $q_{\cap, in}$ for different CdS concentrations

given as parameter

In Fig. 6a, it is worthwhile to note that, for a low catalyst concentration (0,02 g.L$^{-1}$), the mean production rate increases linearly versus incident photon flux density $q_{\cap, in}$ up to a 1700 µmol.m$^{-2}$.s$^{-1}$, for higher PFD the production rate increase is slower approaching the classically observed behaviour where reaction rate is proportional to $q_{\cap, in}^{1/2}$ [36] [37].

At higher concentrations (0.1 and 0.4 g.L$^{-1}$), the apparent linearity cannot be observed in the studied range of photon flux densities, the reaction rates are respectively changing according to $q_{\cap, in}^{0.48}$ and $q_{\cap, in}^{0.53}$ i.e. also according to $q_{\cap, in}^{1/2}$ considering standard deviations.

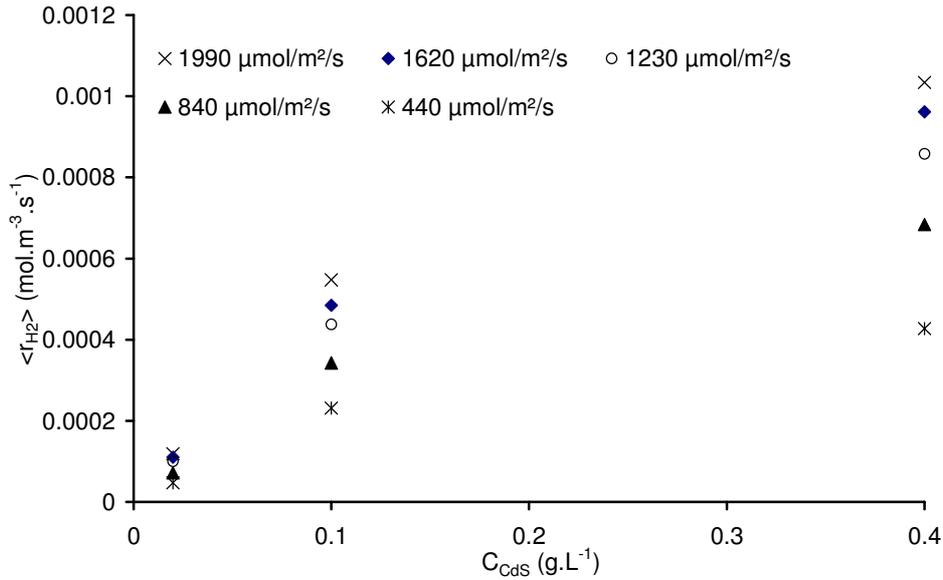

Fig. 6b. Mean production rate of hydrogen versus CdS concentrations for different PFD given as parameter

As it can be seen on Fig. 6b, the production rate increases when catalyst concentration is augmented. Usually in the literature, the reaction rates are as well found to be directly proportional to the catalyst concentration or its mass. This is interpreted as a true heterogeneous catalytic regime [36]. At higher concentrations, the reaction rate levels off and becomes independent of the mass or the concentration [36] or could even decreases [29]. These observations are explained by a screening effect of particles, which masks part of the reactor making it inactive in terms of $H_2$ production.

In the following we propose a radiative analysis of these results, which permits a better understanding of the process.

**4.2 Radiative approach for the description of the results**

Our analysis includes the mean slurry transmittance, $\tau$ (calculated using Eq. 3), that is a radiative quantity independent of the incident PFD $q_{\cap, in}$. Table 1 gathers our measurements of the transmittance as a function of CdS concentration. We recorded constant value of $\tau$ during all the implemented experiments (few hours) for the selected rotation speed of the impeller (at 1100 rpm, there is no evidence to suggest that attrition or sedimentation phenomena occur significantly). $\tau$ is the radiative quantity containing the information on the screening effect of particles (see 4.1). When the photocatalyst concentration increases, screening effect increases (light is more attenuated) and less photons reach the rear of the reactor: the mean transmittance decreases (see Table 1).

| CdS concentration (g.L$^{-1}$) | Mean slurry transmittance $\tau$ (%) | Mean calculated slurry reflection $\rho$ (%) | Mean calculated absorptance $\alpha$ (%) |
|---|---|---|---|
| 0.02 | 80±7 | $\approx 0$ | $20 \pm 7$ |
| 0.1 | 36±3 | 1 | $63 \pm 3$ |
| 0.4 | 6±1 | 3 | $91 \pm 1$ |

Table 1: Mean measured transmittance during $H_2$ photo-production experiments and subsequent estimation of absorptance (see text for calculations of $\alpha$ and $\rho$)

It is possible to notice two operating conditions. For the concentrations of 0.02 and 0.1 g.L$^{-1}$ of CdS, the transmissions measured behind reactor are respectively equal to 80% and 36%. It means that photon absorption is incomplete; a large amount is exiting and thus lost for the photoprocess. At a 0.4 g.L$^{-1}$ catalyst concentration, the attenuation is much more important; the mean transmittance is around 6% only. Nevertheless more than 6% of the photons are lost for the photoprocess. Indeed photons are also lost at the front window due to backscattering within the slurry. The next paragraph focuses on the amount of photons absorbed within the reactor and therefore converted by the photoreaction.

The measurement of $q_{\cap, out}$ is also useful for the estimation of the mean volumetric rate of radiant light energy absorbed (MVREA), $\langle \mathcal{A} \rangle$, *i.e.* the amount of photons absorbed per unit of time and volume that can then be obtained from the energy balance on the photonic phase. This balance can be written as [38]:

$$\langle \mathcal{A} \rangle = -\frac{1}{V} \iiint\limits_{V} (\nabla \cdot \mathbf{q}) \, dV = -\frac{1}{V} \iint\limits_{S} \mathbf{q} \cdot d\mathbf{S} = \frac{1}{V} \left[ S_{in} q_{in} - \sum_{j} S_{out,j} q_{out,j} \right] \tag{7}$$

This relationship is particularly convenient to implement in a flat reactor illuminated on one side that can be considered, in a first approximation, as a one-dimensional infinite slab for photon transport. Moreover, there are only one entrance and one exiting surfaces, $S_{in}$ and $S_{out}$, that are equal in our reactor simplifying Eq. 7. as:

$$\langle \mathcal{A} \rangle = \frac{S_{light}}{V_L} [q_{in} - q_{out}] = a_{light}(q_{in} - q_{out}) \tag{8}$$

Where $q_{in}$ is the net photon flux density at the front windows entering defined as:

$$q_{in} = (1 - \rho) q_{\cap, in} \tag{9}$$

With $\rho$ is the mean slurry reflectance and $a_{light}$ the illuminated specific surface in the rectangular reactor, defined as (Eq. 10):

$$a_{light} = \frac{S_{light}}{V_L} = \frac{S_{light}}{V_{light}} \left(1 - f_d\right) = \frac{1}{L_{reactor}} \left(1 - f_d\right) \tag{10}$$

The maximum mean volumetric rate of radiant light energy, $\langle \mathcal{A}_0 \rangle$, corresponds to the case where no photon is lost neither by transmission nor by reflection in the process:

$$\langle \mathcal{A}_0 \rangle = a_{light} q_{\cap, in} \tag{11}$$

Physical quantities introduced in Eq. 8 and 11 can be used to estimate the global absorptance, $\alpha = \dfrac{\langle \mathcal{A} \rangle}{\langle \mathcal{A}_0 \rangle}$ [39]. Practically, it can be calculated using the transmittance, $\tau$, and reflectance, $\rho$, according to:

$$\frac{\langle \mathcal{A} \rangle}{\langle \mathcal{A}_0 \rangle} = \alpha = 1 - \tau - \rho \tag{12}$$

In other words, $\rho$ is the reflected part of the incoming photons by the medium only. Since this quantity can not be measured in our experimental set-up, we estimated $\rho$ thanks to a two flux model [40] [41] (also refered as the Kubelka-Munk model) using the scattering properties of the CdS particle calculated by Lorenz-Mie theory [42]. $\rho$ values according to CdS concentration are presented in Table 1.

As can be seen in Fig. 6b. the higher $\alpha$ is, the higher $H_2$ production rate: the more photons are absorbed and converted the more $H_2$ is produced. But the maximum value of $\alpha$ is lower than 1 due to slurry reflection [39]: it is not possible to absorb and convert all the photons entering the reactor. In this work the maximum value of $\alpha$ that is investigated corresponds to a 0.4 g.L$^{-1}$ CdS concentration (see Table 1) and is nearly equal to 95% of the maximum possible absorptance $1-\rho$. This approach, based on maximising the absorptance, is different of the analysis developed in [36], considering that the optimal light power utilization corresponds to the domain where the rate is proportional to the PFD, regardless to the presence of dark zones in the reactor or lost photons exiting from the reactor. Now that we analysed absorption within the reactor, we focus hereafter on the conversion of the absorbed photons, with special attention on the apparent quantum yield.

To get further in the analysis of our experimental results, we would like to suggest another approach that doesn't try to relate mean hydrogen production rates $< r_{H_2} >$ to the incident photon flux densities $q_{\cap, in}$, but to the mean volumetric rate of radiant energy absorbed $<\mathcal{A}>$. Such a presentation is more relevant to consider incomplete photon absorption in the reactor [13] [41]. Assuming first a linear kinetic coupling of the $H_2$ production rate with radiation field within the reactor (even to find out later that this is not the case), the photocatalytic rate of $H_2$ production, $< r_{H_2} >$, is:

$$< r_{H_2} > = \Phi \langle \mathcal{A} \rangle \tag{13}$$

Where $\Phi$ is the overall apparent quantum yield (usually considered as independent of the wavelength) and $<\mathcal{A}>$ is the MVREA. This Eq. (14) is justified only if coupling is linear, that is to say, if $\Phi$ is independent of the radiative quantities ($\Phi$ is constant).

Quantum yield values are presented Fig. 7 and are constructed using Eq. 8. (neglecting the rear water-glass interface), 9. and 13. giving the following expression of the quantum yield ([38]):

$$\Phi = \frac{< r_{H_2} >}{a_{light} \left[ (1 - \rho) q_{\cap, in} - q_{\cap, out} \right]} \tag{14}$$

With $< r_{H_2} >$ being determined by pressure evolution, $q_{\cap, in}$ and $q_{\cap, out}$ estimated as explained in section 3.

Nevertheless we record hereafter, in Fig. 7, values of $\Phi$ that vary as a function of the MVREA $<\mathcal{A}>$ and hence $q_{\cap, in}$. This *reductio ad absurdum* proves the non linearity of the coupling between $H_2$ reduction and radiative transfer.

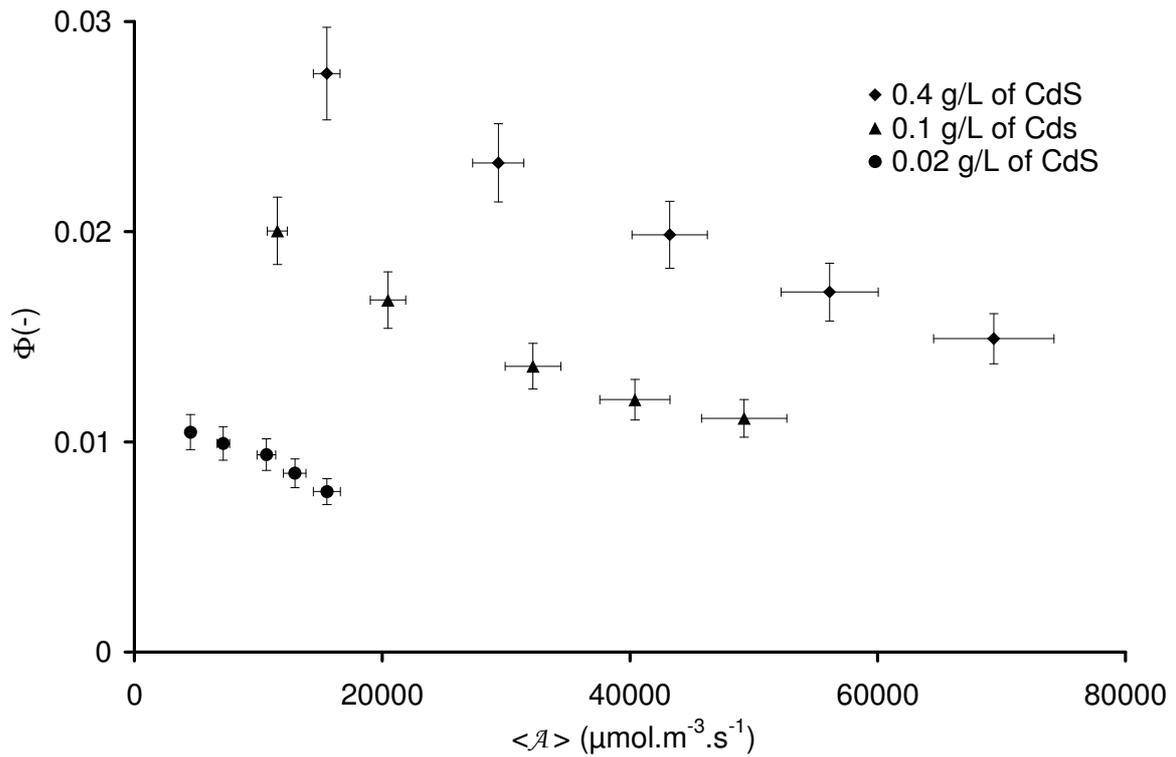

Fig. 7. Evolution of apparent quantum yield, $\Phi$, versus MVREA, $<\mathcal{A}>$, for different concentrations of the photocatalyst

Besides at first glance, the values of $\Phi$ are rather low, a symbol of poor conversion of photon flux into hydrogen without efficient catalyst for redox reactions involved. It should be considered that the immediate goal of this work is not maximizing the quantum yield value through all kinds of chemical activation (Pt [43] or Pd [44] depositions, nitrogen doping [45]…) but rather to develop an experimental bench and a rigorous methodology for accurate radiative analysis of photoreactors before formulating thermo-kinetic coupling in a future work.

Moreover, the "wide" error bars are mainly due to the standard deviation on the measurement of the PFD (about 4%) and not to the method of measuring the hydrogen production rate (standard deviations less than 1 %).

Contrary to our initial simplifying assumption, the apparent quantum yield $\Phi$ clearly decreases when the MVREA increases. This is indeed the direct proof that the coupling is non linear, demonstrating that $\Phi$ depends on the local volumetric rate of radiant energy absorbed, $\mathcal{A}(x)$, that is highly heterogeneous within the reactor operating at high absorptance $\alpha$. This can be explained by a concentration of charge carriers generated in the CdS crystal structure, that increases with local rate of photon absorption $\mathcal{A}(x)$. Different models in the literature (*e.g.* [46]) indicate a first order rate recombination mechanism of these carriers. Increasing their concentrations in the particles is therefore prone to promote the recombination dissipative mechanism (see [47]): the quantum yield is a local quantity that is a function of $\mathcal{A}(x)$.

As can also be seen, the apparent quantum yield augments with catalyst concentration, indirectly confirming the effect of absorptance on the global reactor performances. Using the results presented in Fig. 6a and Eq. 11 and 13, the apparent quantum yield, $\Phi$, could be to, a certain extent, a function of $q_{\cap,\ in}^{-1/2}$ joining observation and data already published [47].

To go further in the analysis of the results, subsequent studies will include the modelling of the local volumetric rate of radiant energy absorbed $\mathcal{A}(x)$, and the formulation of the local thermokinetic coupling, in order to grasp the variation of the quantum yield $\Phi$ as a function of the location within the reactor, along the line of [38] [41] in the field of photobioreactor engineering.

**Conclusions and perspectives**

In this article, a novel characterization bench for the photocatalytic production of $H_2$ was presented comprising three main components, a fully equipped flat torus reactor, a controlled light source (LED panel) and an analysis system. A complete and careful study of these components was carried out. In particular, a study of residence time distribution of the liquid phase in the reactor confirms a piston-like behavior with dispersion and recirculation that tends under certain conditions to a perfectly stirred reactor. Regarding the flux emitted by the LED panel, it has been characterized via two methods: actinometry and hemispherical cosine sensor mapping; the identified photon flux densities with both methods are consistent and accurate.

Both results (perfectly stirred liquid phase and exact knowledge of photon flux entering the reactor) were then used to analyze the process during hydrogen production. Hydrogen was then generated in our device (using CdS and sulphide/sulphite slurry) and its production rate was quantified with a high accuracy thanks to the pressure measurement method. Results were at first presented in a traditional way *i.e.* mean volumetric rates versus incident photon flux and were coherent with literature data.

Subsequently thanks to the estimated production rates combined with macroscopic mass and photonic phase balances, we were able to estimate an apparent quantum yield for radiation conversion into chemical energy. The obtained values were quite low ($\Phi$ of order $10^{-2}$), but as expected since no chemical system optimization was performed on the photocatalyst. Unlike our simplifying assumption about the linear kinetic coupling, $\Phi$ was found to be dependent on the MVREA $<\mathcal{A}>$ for all the tested conditions. Hence a non linear coupling occurs between gas production rate and volumetric rate of radiant energy absorbed.

A more relevant analysis would be to determine the LVREA $\mathcal{A}(x)$ by modeling radiative transfer, this can only be done after solving the radiative transfer equation (RTE) requiring the knowledge of the radiative properties (absorption and scattering coefficients and the phase function) of the catalyst particles. These two difficult points constitutes crucial scientific aspects in their own, or bolts to be unlocked. From these first experimental results presented here and our approach, it will be possible to develop a thermokinetic coupling law and to reify the knowledge models for the specific photocatalytic $H_2$ production. Such models will enable to simulate, design and optimize photoreactors for hydrogen production at further industrial scale demonstrator.

In addition, this installation now validated will also be used to test the effectiveness of new and innovative catalysts. For example, we intend to work with bio-inspired molecules, involved in artificial photosynthesis, either for water reduction (with diiron complexes, exhibiting much higher activity than CdS particles, to replace platinum for the effective reduction of protons, [48]) or organic substrate photo-oxidation [49].

**Acknowledgment**

This work has been sponsored by the French government research program "Investissements d'avenir" through the IMobS3 Laboratory of Excellence (ANR-10-LABX-16-01), by the European Union through the Regional Competitiveness and Employment program -2007-2013- (ERDF – Auvergne region) and by the Auvergne region.

This work was also supported by the National Research Agency through the Tech'Biophyp program (2011-2015). The authors also acknowledge the CNRS research federation FedESol where fruitful discussions and debates take place every year since 2012.

The authors also wish to thank Pascal Lafon and Frederic Joyard for fruitful discussions, CAD and technical assistance.

**Appendix: Full mathematical model of pressure evolution in the gas-tight torus photoreactor for $\langle r_{H_2} \rangle$ assessment**

In this Appendix, we intend to develop a model based on $H_2$ mass balances for the estimation of production rates based on pressure evolution; the global dynamic model (including mean volumetric mass transfer coefficient, $k_L a$) with its simplification (to obtain Eq. 4 in the core of the article) will first be presented. The procedure for estimation of $k_L a$ will subsequently be explained.

**A.1 Global dynamic model for $\langle r_{H_2} \rangle$ assessment**

Following the work of Cogne et al. [26] on the estimation of gas production with pressure increase in a gastight reactor, a complete dynamic model was developed to estimate $\langle r_{H_2} \rangle$ taking into account $H_2$ exchange between liquid and gas phases. For ease of simplification, the contribution of argon (inerting gas used during the reactor preparation) will not be considered in the pressure increase (the experiment begins at equilibrium for inert gas and its partial pressure decreases strongly with time).

The hydrogen balance in the whole reactor (gas and liquid phases) can be described by the following equation (A.1):

$$\frac{d}{dt}\left(C_{H_2,G} V_G + C_{H_2,L} V_L\right) = \langle r_{H_2} \rangle V_L \tag{A.1}$$

Regarding gas phase where pressure is monitored it is possible to write Eq. A.2, considering the mass transfer at the gas/liquid interface:

$$\frac{d}{dt}\left(C_{H_2,G} V_G\right) = k_L a V_L \left(C_{H_2,L} - C_{H_2,L}^{\cdot}\right) \tag{A.2}$$

In equations A.1 and A.2, hydrogen concentrations in liquid and gas phases are appearing. They can be expressed in terms of partial pressures, $P_{H_2}$, the measured parameter:

$$C^{*}_{H_2,L} = \frac{P_{H_2}}{H_{H_2}} \quad (A.3) \; ; \; C_{H_2,G} = \frac{P_{H_2}}{RT} \quad (A.4)$$

In which we have used (Eq. A.3), the Henry's law to relate $H_2$ partial pressure $P_{H_2}$ to the hydrogen equilibrium concentration in the liquid phase $C^{*}_{H_2,L}$. As a reminder, its value is estimated to be 171033 $m^3.Pa.mol^{-1}$ (see text).

Equations (A.1) and (A.2) can then be rewritten as:

$$\frac{d}{dt}\left(\frac{P_{H_2}}{RT} V_G + C_{H_2,L} V_L\right) = \langle r_{H_2}\rangle V_L \qquad (A.5)$$

$$\frac{d}{dt}\left(\frac{P_{H_2}}{RT} V_G\right) = k_L a V_L\left(C_{H_2,L} - \frac{P_{H_2}}{H_{H_2}}\right) \qquad (A.6)$$

Observing that the accumulation term of Eq. (A.5) is the total number of $H_2$ moles on both phases, it is possible to integrate this equation first to obtain a mathematical relation $C_{H_2,L} = f(t, \langle r_{H_2}\rangle, P_{H_2})$ that can be used in Eq. (A.6) to give Eq. (A.7) (where $C_0$ is, for simplicity, defined as the initial hydrogen concentration in liquid phase):

$$\frac{dP_{H_2}}{dt} + k_L aRT \frac{V_L}{V_G}\left(\frac{1}{RT}\frac{V_G}{V_L} + \frac{1}{H_{H_2}}\right)P_{H_2} = k_L aRT \frac{V_L}{V_G}\langle r_{H_2}\rangle t + k_L aRT \frac{V_L}{V_G}\left(C_0 + \frac{P_0}{RT}\frac{V_G}{V_L}\right) \qquad (A.7)$$

The solution of this linear ordinary differential equation with $P_{H_2}(t=0) = P_0$ is:

$$P_{H_2} = \frac{k_L aRT}{\chi}\frac{V_L}{V_G}\langle r_{H_2}\rangle t + \frac{k_L a}{\chi}\left(RT\frac{V_L}{V_G}\left(C_0 - \frac{\langle r_{H_2}\rangle}{\chi}\right) + P_0\right)\left(1 - e^{-\chi t}\right) + P_0 e^{-\chi t} \qquad (A.8)$$

Where for ease of reading the $\chi$ parameter, having a constant value during experiment, has been introduced (Eq. A.9):

$$\chi = k_L aRT \, \frac{V_L}{V_G} \left( \frac{1}{RT} \frac{V_G}{V_L} + \frac{1}{H_{H_2}} \right) \tag{A.9}$$

As already explained, the estimation of $\langle r_{H_2} \rangle$ is based on the measurement of pressure evolution in the reactor as a function of time $\frac{dP_{H_2}}{dt}$. It is thus necessary to differentiate Eq. (A.8) to obtain the useful relation (A.10):

$$\frac{dP_{H_2}}{dt} = \frac{k_L aRT}{\chi} \frac{V_L}{V_G} \langle r_{H_2} \rangle + k_L a \left( RT \frac{V_L}{V_G} \left( C_0 - \frac{\langle r_{H_2} \rangle}{\chi} \right) + P_0 \right) e^{-\chi t} - \chi P_0 e^{-\chi t} \tag{A.10}$$

Eq. (A.10) is composed of two terms, a steady one and another time-dependent. For 'sufficiently long times' (see hereafter), Eq. (A.10) can be simplified as:

$$\xrightarrow{t \to \infty} \quad \frac{dP_{H_2}}{dt} = \frac{k_L aRT}{\chi} \frac{V_L}{V_G} \langle r_{H_2} \rangle \tag{A.11}$$

$$\Leftrightarrow \langle r_{H_2} \rangle = \frac{\chi}{k_L aRT} \frac{V_G}{V_L} \frac{dP_{H_2}}{dt} \tag{A.12}$$

Considering the expression of $\chi$ (Eq. (A.9)), mean volumetric rate of hydrogen production can then be determined without the need of considering gas/liquid mass transfer as proposed in the core of the article (Eq. 4).

$$\langle r_{H_2} \rangle \overset{t \to \infty}{=} \left( \frac{1}{RT} \frac{V_G}{V_L} + \frac{1}{H_{H_2}} \right) \frac{dP_{H_2}}{dt} \tag{A.13}$$

## A.2 Determination of the time constant to neglect mass transfer in the assessment of $\langle r_{H_2} \rangle$

However, the expression 'sufficiently long times' should be stated. To respond to this requirement, Eq. A.10 must be considered and mean volumetric mass transfer coefficient, $k_L a$, and initial concentration, $C_0$, need to be estimated.

For this purpose, the experimental procedure is as follows: after hours of hydrogen production, the incident photon flux density is set at a zero value by switching off the LEDs and covering the reactor. Hydrogen production would cease due to the absence of irradiation. Besides, pressure in the reactor is still monitored and increases as a function of time, as can be seen in Fig. A.1. This evolution is due to $H_2$ transfer from the liquid phase to the gas phase tending to the thermodynamic equilibrium.

First the value of $C_0$ is determined by integral mass balance on $H_2$ for the duration of the experiment at dark according to Eq. A14:

$$C_0 = \frac{P_{H_2,\infty}}{H_{H_2}} + \frac{V_G}{RTV_L}\left(P_{H_2,\infty} - P_{H_2,0}\right) \qquad (A.14)$$

Secondly, the value of $k_La$ is then identified by comparing model in Eq. (A.8) with $\langle r_{H_2} \rangle = 0$ and experimental results as it can be seen for example in Fig. A.1.

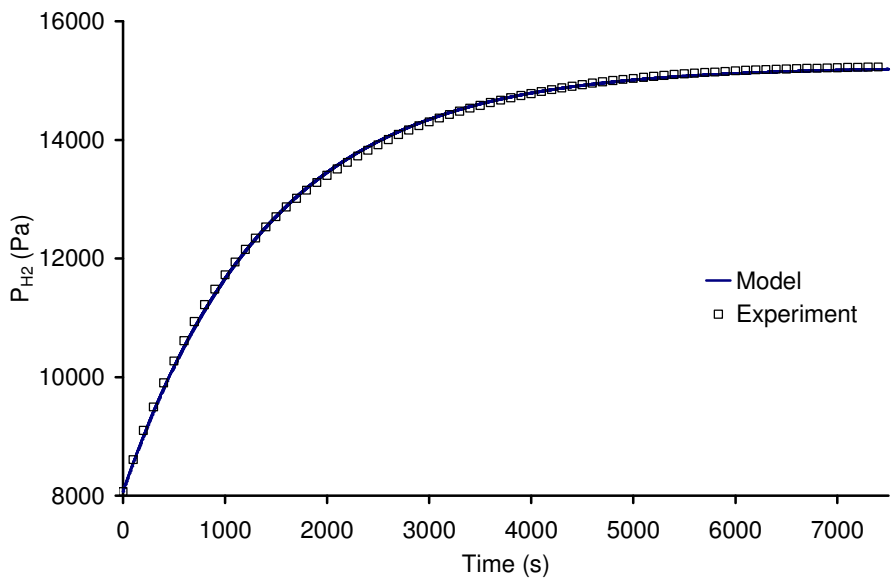

Fig. A.1. Typical pressure evolution at dark after reaction, comparison with model in eq. A.8

after $k_La$ identification

Matching experimental results and model is very good as can be noticed in Fig. (A.1). The identified value of $k_L a$, in this case, is equal to 2.4 h$^{-1}$, in the same order of magnitude as typical volumetric mass transfer coefficients in still surface.

To check the validity of using (Eq. A.13) to estimate the volumetric rate of hydrogen production, a comparison is made between the linear model (Eq. A.11) and the complete model (Eq. A.10) for pressure evolution based on the experimental bench parameters (including the identified value of $k_L a$ and a given $\left\langle r_{H_2} \right\rangle$ ). The results are presented in Fig. A.2.

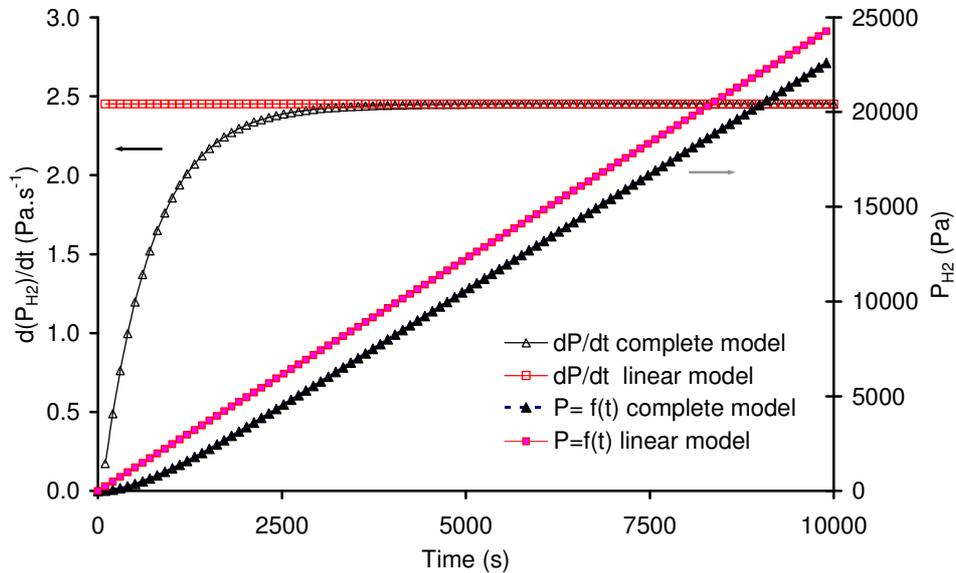

Fig. A.2 Comparisons of pressure evolutions (P=f(t)) and pressure evolution rates (dP/dt=f(t)) respectively predicted by linear and complete models

Due to the existence of liquid/gas transfer phenomenon, pressure evolution predicted by the complete model is not fully linear in the first 3500 seconds. This duration should thus be waited to consider steady state being reached (less than 1% difference between estimated values in pressure evolution rates). Such a precaution was taken into account to make good use of the measured pressures. An example of pressure recording can be seen in Fig. A.3, demonstrating a very good agreement with previous theoretical analysis of the time constant. The mean volumetric rates of hydrogen evolution are then estimated from the linear part.

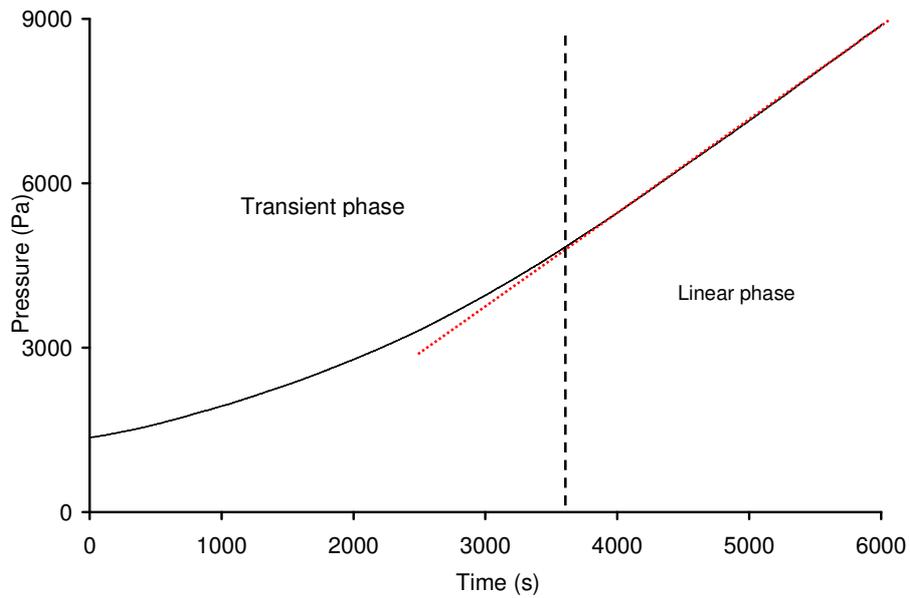

Fig. A.3. Example of experimental pressure recording presenting transient and linear phases

**Nomenclature**

| | |
|---|---|
| $\mathscr{A}$ | Local volumetric rate of radiant energy absorbed (LVREA) ($\mu$mol.m$^{-3}$.s$^{-1}$) |
| $\mathscr{A}_0$ | Maximum local volumetric rate of radiant energy absorbed ($\mu$mol.m$^{-3}$.s$^{-1}$) |
| a | Specific surface of liquid/gas transfer (m$^{-1}$) |
| $a_{light}$ | Illuminated specific surface (m$^{-1}$) |
| C | Concentration (mol.m$^{-3}$) |
| $C_\infty$ | Final salt concentration in the reactor (mol.m$^{-3}$) |
| $C^*$ | Liquid concentration at gas/liquid equilibrium (mol.m$^{-3}$) |
| $D_{ax}$ | Axial dispersion coefficient (m$^2$.s$^{-1}$) |
| $f_d$ | Dark fraction of the reactor (-) |
| H | Henry constant (Pa.m$^3$.mol$^{-1}$) |
| j | Index of summation (-) |
| $k_L$ | Liquid/gas mass transfer coefficient (m.s$^{-1}$) |
| L | Characteristic length of the flow (m) |
| $L_{reactor}$ | Thickness of the reactor (m) |
| P | Pressure in the reactor (Pa) |
| q | Photon flux density ($\mu$mol.m$^{-2}$.s$^{-1}$) |
| R | Gas constant (8,314 J.mol$^{-1}$.K$^{-1}$) |
| $r_i$ | Local molar volumetric reaction rate (mol.m$^{-3}$.s$^{-1}$) |
| S | Surface (m$^2$) |
| $S_{light}$ | Illuminated surface (m$^2$) |
| T | Temperature (K) |
| t | Time (s) |
| $t_c$ | Circulation time (s) |
| $t_m$ | Mixing time (s) |
| v | Liquid mean velocity (m.s$^{-1}$) |
| $V_G$ | Headspace gas volume in the reactor (m$^3$) |
| $V_L$ | Volume of solution contained in the reactor (m$^3$) |
| $V_{light}$ | Illuminated volume of the reactor (m$^3$) |
| x | Space coordinate in the Cartesian geometry (-) |
| z* | Reduced distance between tracer injection and measurement (-) |

**Greek letters**

$\alpha$             Absorptance (-)

$\chi$             Time constant in the complete model for pressure evolution ($s^{-1}$)

$\Phi$             Apparent overall quantum yield ($mol_{H2}.mole_{h\nu}^{-1}$)

$\Theta$             Dimensionless time $\Theta = \dfrac{t}{t_c}$ (-)

$\lambda$             Wavelength (m)

$\rho$             Reflected part of the incoming photons by the medium itself (-)

$\tau$             Mean transmittance of the slurry (-)

$T$             Mean transmittance of the window glass

**Dimensionless number**

Pe             Péclet number based on the length of the reactor (-) $Pe = \dfrac{v \cdot L}{D}$

**Subscript**

$\cap$             Relative to hemispherical incident radiation onto the reactor

$\lambda$             Relative to a spectral quantity for the wavelength

$\infty$             Relative to a long-term time

0             Initial value

G             Relative to gas phase

gas             Relative to gaseous solute

glass             Relative to the window glass

$H_2$             Relative to $H_2$

in             Relative to inlet

L             Relative to liquid phase

out uncorrected Relative to outlet without interface correction

out             Relative to outlet

**Other**

$\langle X \rangle = \dfrac{1}{V} \iiint_V X dV :$      Spatial averaging